# Agent-Based Simulation of a Perpetual Futures Market

*Ramshreyas Rao*

## Abstract

I introduce an agent-based model of a Perpetual Futures market with heterogeneous agents trading via a central-limit order book. Perpetual Futures (henceforth Perps) are financial derivatives introduced by the economist Robert Shiller, designed to 'peg' their price to that of the underlying Spot market. This paper extends the limit order book model of Chiarella et al. (2002) by taking their agent and orderbook parameters, designed for a simple stock exchange, and applying it to the more complex environment of a Perp market with long and short traders who exhibit both positional and basis-trading behaviors. I find that despite the simplicity of the agent behavior, the simulation is able to reproduce the most salient feature of a Perp market, the 'pegging' of the Perp price to the underlying Spot price. In contrast to fundamental simulations of stock markets which aim to reproduce empirically observed stylized facts such as the leptokurtosis and heteroscedasticity of returns, volatility clustering and others, in derivatives markets many of these features are provided exogenously by the underlying Spot price signal. This is especially true of Perps since the derivative is designed to mimic the price of the Spot market. Therefore, this paper will focus exclusively on analyzing how market and agent parameters such as order lifetime, trading horizon and spread affect the premiums at which Perps trade with respect to the underlying Spot market. I show that this simulation provides a simple and robust environment for exploring the dynamics of Perpetual Futures markets and their microstructure in this regard. Lastly, I explore the ability of the model to reproduce the effects of biasing long traders to trade positionally and short traders to basis-trade, which was the original intention behind the market design, and is a tendency observed empirically in real Perp markets.

# Table of Contents

# 1. Introduction

Perpetual Futures are a type of financial derivative introduced by the Nobel Prize winning economist Robert Shiller (1993), to provide a mechanism of price discovery for assets that may be illiquid, or whose value is difficult to measure directly. While largely ignored in conventional finance, this derivative has found large-scale adoption in cryptocurrency markets, beginning with the introduction of Perpetual Futures by the BitMEX exchange in 2016. Today, the volumes of crypto Perp markets typically dwarf the volumes of the underlying 'spot' markets they are based on (Baur & Dimpfl, 2019).

Perps are designed to 'peg' the price of an underlying asset (the 'Spot' market), which means that their prices evolve in the same pattern as the underlying asset. For example, a Bitcoin Perpetual Futures market will track the prices of a Bitcoin 'Spot' market (cryptocurrency exchanges where Bitcoin is bought and sold). This peg is achieved through a periodic 'funding mechanism' that rewards traders based on whether they go 'long' or 'short' when they enter the market. At the time of funding, if the price in the futures market is higher than the price of the reference asset, the 'long' traders pay the 'shorts' a 'funding rate' based on the asset price, and vice versa. This occurs perpetually at a fixed periodicity.

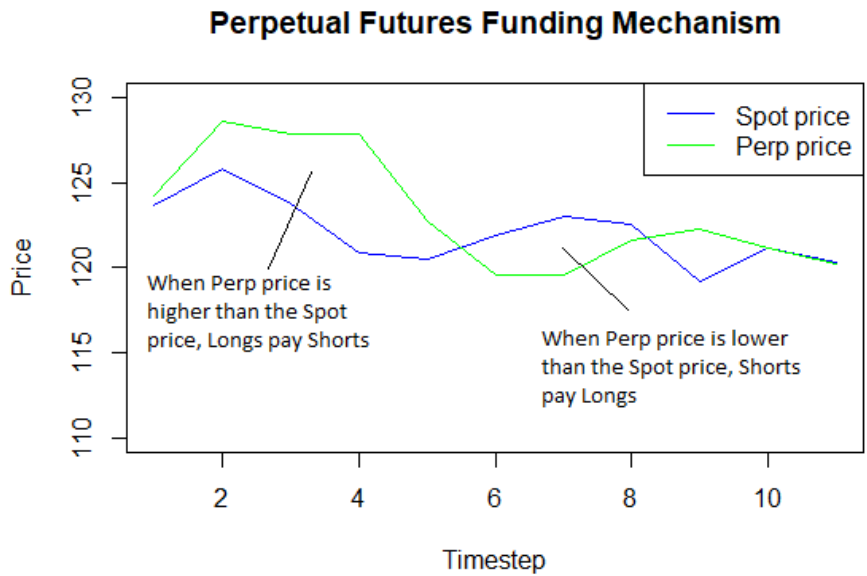

*Figure 1: When the Perp is trading at a premium to the Spot price, the Longs pay the Shorts a 'funding rate' (and vice-verse) based on a formula that may vary from venue to venue. The funding rate is totalled and paid out in fixed intervals. This means whenever the Perp price deviates from the Spot price, market pressure can bring the price back into the 'Peg'.*

The incentive to receive the funding rate by selecting an appropriate bid or ask is sufficient to 'peg' the price of the Perpetual Future to the price of the actual asset even during the interval between funding events. If the price in the Perpetual Future market is higher than the Spot price (trading at premium), the 'longs' pay the 'shorts' the funding rate. This attracts more shorts to the market, exerting a downward pressure to bring the Perpetual price back in line with the index. Conversely, if the Perpetual trades below the Spot price, the shorts pay the longs the same funding rate, attracting longs to the market and exerting an upward pressure to return the Perpetual price to the index. This is an emergent effect of multiple traders attempting to capture the funding payments and has been borne out empirically (Baur & Dimpfl, 2019).

It is important to note that not all traders trade for the funding rate, however. Many traders also trade positionally, which is to buy low, sell high if in a long position, and sell high, buy low if in a short position. This is especially the case due to the prevalence of margin trading in Perpetual Futures exchanges. Since the Perp price tracks that of the underlying asset, traders can use Perps as a proxy for the Spot asset, and trade with leverage for greater returns. The dynamics of the Perp market therefore reflects all these trading behaviors, providing for a rich environment where traders can open both long and short positions, basis-trade for funding payments, or trade positionally.

While the literature is replete with studies of generic stock markets and their derivatives, little research has been done on Perps, and none from an Agent-Based Modeling perspective to my knowledge. In this paper we extend the literature by modeling the Perp market with modified versions of canonical Agent-Based Models of financial markets. The purpose of this paper is to (1) estimate agent parameters that optimally reflect empirical data, and (2) establish a robust framework for performing experiments using the model.

## 2. Literature review

### 2.1 Perpetual Futures

Robert Shiller introduced the Perpetual Future in a presentation to the American Finance Association Meetings in 1993. The purpose of the derivative was to provide a market for hedging assets that are critical to economies such as human capital, real estate, consumer price indices and other crucial economic indicators. He proposes that there are two separate but related measurement problems that make this difficult. One, the prices of these assets may only be observed infrequently, and on heterogeneous types of such assets (infrequent and diverse house sales, for example) and therefore serve poorly as price signals for the entire asset class. The second problem he suggests is that for many assets, there is no price signal at all, but only measurements of the dividend or rent of an asset (salaries for human labor). To address these two problems, Shiller proposes a synthetic derivative marketplace where a 'hedonic repeated price index' is supplied to a market for perpetual claims on cash flows, paid from long traders to short traders, which represent the dividends or rents on these assets. The claims are perpetual to capture the stream of dividends accruing to an asset over its entire existence, therefore enabling price discovery. So, a hedonic, repeated price index would supply a stream of price signals to the perpetual claims market, where traders with long positions would pay a cash-settled dividend to traders who take an equal and opposite short position. This is done in such a way that the price of the perpetual claim tracks the price of the underlying, as described in the introduction and Figure 1.

Given their novelty, at least in practice, there are few studies focusing on Perps. Alexander et al. (2019) find that the prices in the Perp markets lead the prices in most spot exchanges, suggesting that price discovery for the underlying asset in fact occurs in the Perp market, rather than the other way around. Deng et al (2019, 2020) describe optimal trading as well as hedging strategies for these derivatives. Nimmagadda and Sasanka (2019) demonstrate the heteroskedasticity of the funding rate and suggest a causal relation between the funding rate and price of the Perps. Wu (2020) uses regression, stochastic calculus, and numerical simulation methods to provide a quantitative description of the BitMEX XBTCUSD inverse perpetual contract.

**2.2 Agent-based Computational Economics**

While analytical models of complex systems such as stock exchanges have had success describing the aggregate evolution of parameters such as price, they often have little to say about market microstructure and the diversity of agents that populate them (Pagan 1996). Moreover, empirically observed 'stylized facts' such as volatility clustering, leptokurtosis and heteroskedasticity of returns are poorly explained by analytical models (Pagan 1996). Computational simulation in general, and Agent-Based Modeling (ABM) in particular, have had success in reproducing these 'stylized facts', lending explanatory power to the rules governing the simulated agents and their environment (Cramer and Trimborn, 2019).

The application of ABM to financial markets goes back to before the 1990s, applied by Kim and Markowitz (1989) to explain the 'Flash Crash' of 1987 - an anomaly that was unexplained by conventional economic theory. The Santa Fe Artificial Stock Market (Arthur et al. 1997) was another notable early Agent-Based model, along with work by Levy, Levy and Solomon (1994). Numerous developments since these early beginnings have grown the field to what is now referred to as Agent-based Computational Economics (Tesfatsion, 2003). An extensive survey of the field with motivations for its application to problems that analytic approaches cannot solve is in LeBaron (2006). Chen, Chang, and Du (2012) describe a 'three-stage development' of Agent-based Computational Economics (ACE) in a comprehensive review of the field. In particular, they frame the three stages in terms of their relationship to the field of econometrics, which does not look at economic systems as complex adaptive systems, but rather as statistical models of data-generating processes inspired by models of gasses and other ensembles in Physics (Arthur, Durlauf & Lane, 1997). I use this framing to review the literature regarding the application of ACE to financial markets below.

**2.3 Presenting ACE with Econometrics**

The first stage in the development of ACE is "presenting ACE with econometrics". Here, the data generated by an ACE model is analyzed to determine if they reproduce the 'stylized facts' mentioned above. The focus is on agent design, which ranges from 'few-type' models with less than three agent types, to n-type models with an arbitrary number of types - also called 'many-type' models.

An early, seminal work by Gode and Sunder (1993) uses a parsimonious design of a single agent - a 'Zero Intelligence trader', which is surprisingly able to reproduce the stylized facts. This work served as a foundation for many future works, further developing agent designs. Gaunersdorfer and Hommes (2007) detailed an early model of a stock market with two types of traders, 'Fundamentalist', who believe the stock has a true value function which the price is expected to revert to, and 'Trend following', who believe that past trends tend to continue into the future. An earlier work by Farmer and Joshi (2001) adds another trader type, 'market makers', who absorb fluctuations in excess demand, lowering the price when they must buy and raising it when they must sell. These 'few-type' models laid the groundwork for the design of ACE models that reproduce the empirically observed stylized facts.

**2.4 Building ACE with econometrics**

The second stage is "Building ACE with econometrics". In this stage, the question of estimating the best model parameters for the closest fit to empirical data is explored. Franke and Westerhoff (2012) extend the previous papers by enabling the distribution of these agent types to evolve towards better performing strategies over multiple runs and describe the best fit to experimental data. Boswijk, Hommes and Manzan (2006) use historical stock price data to act as the true value function, and evolving ratios of fundamentalist and trend following traders based on recently successful strategies. This paper finds 'expectation regimes' where the market enters a regime of majority trend followers driving prices beyond fundamental value, as well as 'fundamentalist' regimes when the stock prices cleave closely to the expected fundamental value, as was observed in the historical data. LeBaron and Yamamoto (2006) also use learning agents to show that the model is able to replicate long-memories in trading volume, volatility in returns, and signs of market orders.

**2.5 Emerging econometrics with ACE**

The third and last stage is "Emerging econometrics with ACE". This is a nascent field, where agent-based models are used to make novel contributions or examine econometric theory. For example, the aggregation problem, which questions the validity of aggregating quantities such as demand over diverse individuals (Stoker, 1993). Chen, Huang and Wang (2008) use an agent-based Capital Asset Pricing Model (Sharpe, 1964) to show that the aggregation assumption (of agents being identical) leads to incorrect results.

By creating a data-generating process where the internals are visible to researchers, unlike analytic approaches, ACE can be used to examine the simplifications necessary for current econometric theory. Apart from this, agent-based models can also be used to make contributions in industry applications, as Alexander, Deng, and Zou (2021) did with their paper detailing optimal hedging strategies using Bitcoin Futures.

The agents and models described above therefore share a common lineage that leads back to a small set of original designs. These canonical agents serve as starting points for the design of many ACE models. This brief review of ACE serves the purpose of motivating the structure for the paper. As a novel agent-based model of a particular type of financial market, the primary goal of "Presenting the ACE model with econometrics" will be the first objective, i.e., to reproduce empirically observed stylized facts - in this case the 'peg' of the Perp price to the Spot price. This is done by applying the canonical agents described in the work of Chiarella et al. (2002) to a simulated Perpetual Futures market.

**2.6 A simulated analysis of the microstructure of double auction markets, Chiarella et al. (2002)**

The aim of the paper is to 'introduce a modeling framework for the analysis of an order-driven market based on numerical simulations.' Chiarella et al. (2002) aimed to create a simple model of a limit-order book exchange while still preserving the essential structure of the market. One of their purposes was to provide it with the ability to be 'embellished' to enable the study of the effect of market protocols and rules on market behavior.

Chiarella et al. (2002) define agents as a randomly weighted composites of Fundamental, Chartist and Noise traders. Each strategy is associated with a forecasting function that takes the current price and produces a price forecast. The three forecasts are combined according to random weights to produce a composite forecast. If the forecast is greater than the current price, they submit a Bid (buy order), otherwise the agent submits an Ask (sell order). Like many simulations, including my own in this paper, agent wealth is not a consideration, as market dynamics rather than trading strategy are the main concern. At any time t, an agent is randomly selected to participate in trading. The price is set by a successful trade (bids and asks match), or the mid-point of the orderbook if no trades occur, as is the case in my model.

The authors find that their model captures real world factors such as spread of trader behavior represented by the Fundamentalist, Chartist and Noise traders, and identify parameters that represent key determinants of order flow dynamics such as tick size, liquidity, and the average lifetime of a market order. The model is presented as a foundation on which to build future enquiries such as relaxing the Fundamentalist forecast, which assumes the stock has a fundamental value known to investors (which I do in this paper), as well as taking trader budget constraints into account and incorporating dynamic order sizing (both of which I do not). As a robust simulation with rich dynamics utilizing simple agents, this model represents a suitably capable foundation upon which to develop the much more complex environment of a Perp market.

## 3. Methods

The simulation code uses the Python implementation of the Chiarella et al. paper from the AgentFin website by Blake LeBaron of Brandeis University as a foundation, which I reimplemented and then extended in R (2016).

### 3.1 Agents

In Chiarella et al. (2002) (henceforth, the original work), the agents are randomly weighted composites of three different types of traders, Fundamentalists, who believe the true value of the asset is a constant value that the price must revert to, Chartists, who use prices within a historical time horizon to produce forecasts, and Noise traders, who produce a random forecast based on a constant volatility parameter and the current price. The three forecasts are combined according to the weights of each trader type to produce a composite forecast. Prior to simulation, a pool of traders is generated with a randomly generated weightage of the three types described above for each trader. Every timestep, a single trader is chosen from this pool to enter the market.

### 3.2 Trading Cohort

This leads to the first variation of my simulation from the original. In the original work, trader entry was probabilistic, and represented liquidity. In my work, a fixed cohort size of randomly drawn traders enters the market at each timestep. The size of the cohort entering at each round of trading is controlled by a constant parameter. This parameter thereby controls the depth of the orderbook, which affects the efficiency of an exchange. While the effect of random entry and cohort size is a very interesting parameter to explore in terms of the effects of liquidity, this exploration was left out of the current paper in the interest of scope and time limitations. A cohort size of four was fixed for all the simulation runs in this paper.

**3.2 Long and Short positions**

As a model of a simple stock market, in the original work traders were not explicitly designed to have long or short positions, though it was possible for the Chartists and Noise traders to have negative weightage, reflecting 'contrarian' strategies. This means a Chartist may produce a negative forecast even if the prices are rising in its time horizon. Such a trader may produce an Ask (sell order) in response to rising prices, similar to the behavior of a short trader. However, the agent still places Bids when their forecast is higher than the current price, and Asks when the forecast is lower, consistent with Long positions.

The second aspect in which my work diverges from the original is the explicit assignment of Long and Short positions to traders when they enter the market. This is more relevant to the Perp exchange, where a trader can only open a Long position if a counterparty takes a Short position on the other side of the trade, and where funding payments depend on the side of the trader. This is unlike a simple stock market where both the buyer and seller are aiming to buy low and sell high. In my simulation, traders start off without an explicit position, and are randomly assigned with equal probability to either Long or Short positions when they enter the market. Once they are assigned a Long or Short position, they trade accordingly for the rest of their time in the market. This is important because 'Longs' pay 'Shorts' if the Perp is trading at a premium, and vice-versa - making it essential to tag traders accordingly and track their positions as long as they are in the market.

**3.3 Close Positions and Exit**

This leads to another variation in my simulation from the original. Every timestep, traders are randomly chosen to exit the market, which is equivalent to them closing their positions, and becoming 'neutral' - as they were before entering the market. This enables them to re-enter the market on the opposite side in a later timestep. This is done so that the population of traders does not end up locked over time into two groups of Longs and Shorts once all of them enter the market, with their particular distributions of parameters then biasing all subsequent rounds of trading. A skew in these distributions can end up producing effects which are artifacts of the distributions, rather than characteristics of a Perp market in general. Randomly closing positions and allowing reentry enables the distribution of the two sides to vary randomly over the entire duration of the simulation, removing such artifacts.

**3.4 Positional Trading vs Basis trading**

The most important extension to the original work is the ability of traders to both basis-trade as well as trade positionally. Positional traders, whether long or short, are trading on the price action (buy low, sell high for Longs, or the reverse for Shorts). Basis traders, in contrast, are trading for the 'Premium' - the difference between the Perp and Spot prices. As described above, if the Perp market is trading at a premium, the 'Longs' periodically pay the 'Shorts' a funding rate. A trader targeting this funding rate may be indifferent to the price action of the two signals in isolation, earning profits as long as one signal consistently trades at a premium to the other (if Bitcoin Perps keep trading at a premium to the Bitcoin Spot market, Shorts in the Perp market keep earning the funding rate). One compelling, and common strategy in the finance industry is to hedge long positions in an asset by taking a short position in the corresponding Perp market to produce a Delta-neutral position that generates returns no matter which way the market is trending, provided the Perp market consistently trades at a premium. Empirically, it is observed that Long traders tend to trade positionally, and Shorts tend to basis-trade - an artifact of the hedging strategy mentioned above, and the presence of leverage in Futures markets. This is controlled by the bias parameter in my simulation, which moderates the propensity for traders to basis-trade or trade positionally at each timestep based on whether they are Long or Short. This is discussed further in the results section.

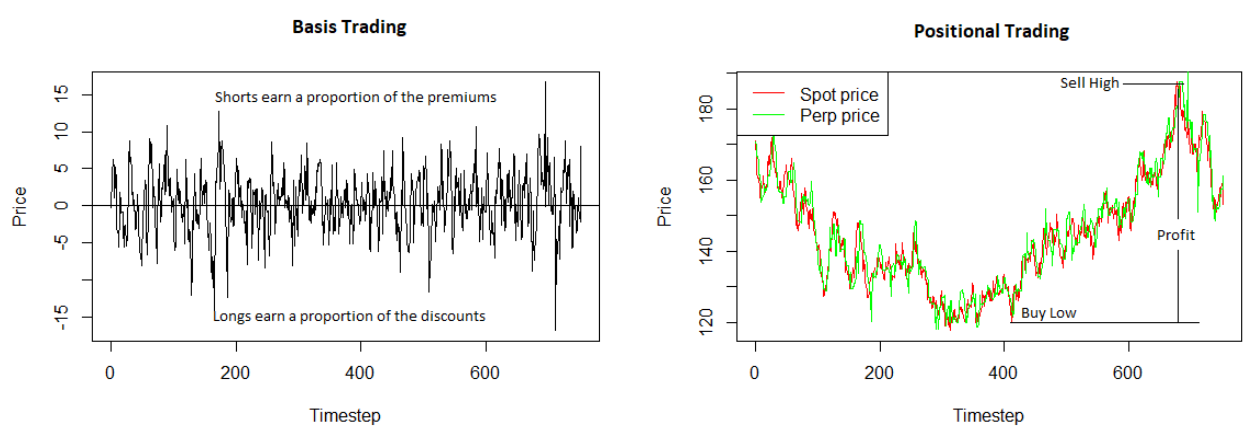

*Figure 2: Positional traders Buy/Sell Low/High as seen on the left, while basis-traders trade for funding rates generated by earning a proportion of the premiums/discounts as seen on the right.*

Once the trading style is established, the trader uses exactly the same machinery implemented in the original work to produce forecasts – the only difference is that the positional trader uses the historical Spot signal to produce forecasts, and the basis-trader uses the historical Premium to forecast future premiums. In this way, I add the required complexity of funding rates to the simulation whilst remaining faithful to the Agent behavior of the original work.

**3.5 Fundamentalist Trader**

While this trader strategy was relevant to the original work as the price evolution was endogenously determined by trader behavior in the simulation, in this simulation there is an external Spot price signal which is the primary driver of value in the Perp market. The notion of a pre-existing, endogenous and constant 'true value' of the asset does not make sense for a Perp simulation which is designed to 'peg' prices to a Spot price signal. Therefore, the weightage of this trader strategy is set to zero in this simulation. The fundamentalist trader aspect does not participate in the simulation at all.

### 3.6 Chartist Trader

The Chartist trader in the original work used a moving average for a time horizon randomly drawn from a range ($l_{min}$, $l_{max}$) of the prices to produce a forecast in each assay. Since there was only one, endogenously produced price history in this simple model, the price history of the simulation itself was used to produce the forecast. In my Perp model, however, there are three signals, the Spot price from the underlying market, the prices evolving in the Perp market itself, and the Premiums, which are defined as the difference between the two. In my Perp model, the Chartist trader uses a moving average (randomly drawn from the time horizon range ($l_{min}$, $l_{max}$)) of the Spot market to produce a forecast for the Perp price if they are trading positionally, and a moving average of the Premiums if they are basis-trading. The Chartist forecast return is computed as follows:

$$r_t = \frac{1}{L} \sum_{j=1}^{L} \frac{p_{t-j} - p_{t-j-1}}{p_{t-j-1}}$$

Where L is drawn from ($l_{min}$, $l_{max}$)

### 3.7 Noise Trader

Similarly, the Noise trader uses a volatility parameter to produce forecasts of the price, and places Bids or Asks depending on whether they are Long or Short. Like the Chartist, they use either the Spot price or the Premium to produce the forecast depending on whether they are trading positionally or basis-trading using the following formula:

$$r_t = \sigma_\epsilon U(0, 1)$$

Where U is randomly drawn from the interval (0, 1).

### 3.8 Composite forecast

The forecasts produced according to the rules above are composed together to form a composite forecast for each trader, using the following formula:

$$r_t = \frac{1}{w_c + w_n} [w_c f_c + w_n f_n]$$

Where $w_c$, $w_n$ are the weights of the Chartist and Noise strategies, and $f_c$, $f_n$ are the Chartist and Noise forecasts as defined above. Here, the forecast is made for the next time step, unlike the original work, where the forecast is made for a period τ. This forecasted return is then used to create a price forecast for the next timestep using the following formula:

$$p_{t+1} = p_t e^{r_t}$$

In the original work, any orders older than τ were also cleared from the orderbook (discussed below). In my simulation, these two concerns are separate, with traders making forecasts for the next timestep, and τ being used solely to remove old orders. If the forecast is greater than the current Price/Premium, the Long/Short trader posts a Bid/Ask according to the following formula:

$$bid_t = p_t(1 - k)$$
$$ask_t = p_t(1 + k)$$

Where $k$ is drawn from ($k_{min}$, $k_{max}$).

### 3.9 τ

In the original work, all orders older than $t - τ$ were cleared from the orderbook, since forecasts were produced for τ *periods into the future*. In my simulation, I use τ differently. Only the last τ orders in the orderbook are retained at each time step. So if there are 10 orders in the orderbook, and τ = 4, only the 4 latest orders (whether they are Bids or Asks) are retained, and the rest removed.

### 3.10 Geometric Brownian Motion for the Spot market signal

Unlike the original work, my simulation requires a Spot price signal that can be repeatedly generated for multiple simulation runs. For simplicity, I use a Geometric Brownian Motion (GBM) for this signal, enabling me to produce multiple signals with the same set of parameters.

### 3.11 Simulation Runs

For each run, a GBM is first generated. Then, a pool of agents is generated based on the parameters we want to investigate. Finally, a hundred simulations are run for each value of the parameter that we are investigating. For example, if we are seeing the effect of varying τ from 2 *to* 12, the GBM and pool of traders (both of whose parameters are fixed) are generated and the simulation is run a hundred times for each value of τ in the range (2, 12). The point estimates of interest are measured and averaged across the hundred runs and then analyzed. This ensures that a chance distribution of a particular generation of traders or the GBM does not skew the results.

Since there are many parameters to explore, to avoid a combinatorial explosion of parameter sweeps some parameters are fixed after a preliminary (and naïve) exploration of the simulation behavior in single runs. Particularly, the relative weightage of the agent strategies is set to an equal weight prior to running the primary simulations. This is discussed below in the results section.

## 4. Simulations and Results

We first consider a Spot market whose prices are represented by a Geometric Brownian Motion:

$$\Delta S_t = \mu \Delta t + \sigma \epsilon \sqrt{\Delta t}$$

Here $S_t$ is the stock price at timestep $t$, μ is a constant 'drift' parameter, and σ is a constant volatility parameter. $\Delta S_t$ is the increment or decrement applied to the current spot price at time $t$ to generate the spot price at $t + 1$.

The *Premium* at any time $t$ is defined as:

$$Premium_t = Perp\ price_t - Spot\ price_t$$

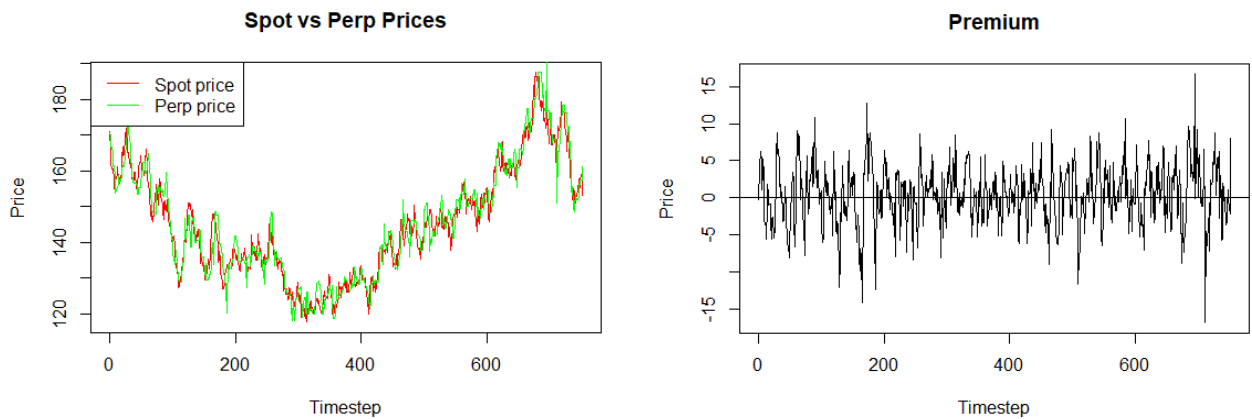

*Figure 3: Results of a simulation run with μ=1, σ=0.5 and S0=100. The 'peg' is clearly holding, with the Perp prices in green tracking the spot prices in red. On the right we see a chart of the premium, the difference between the Perp and Spot prices over time.*

The figure on the left above demonstrates the 'pegging' of the Perp price with the Spot price in a single simulation run. The figure on the right shows the premiums, with a mean that appears close to zero. To see if there is a lag in 'catching up' with the Spot price signal, we use a Pearson's cross-correlation between the Perp and Spot prices to see that the two signals are highly correlated, with a maximum correlation at a lag of 2 for the sample simulation run.

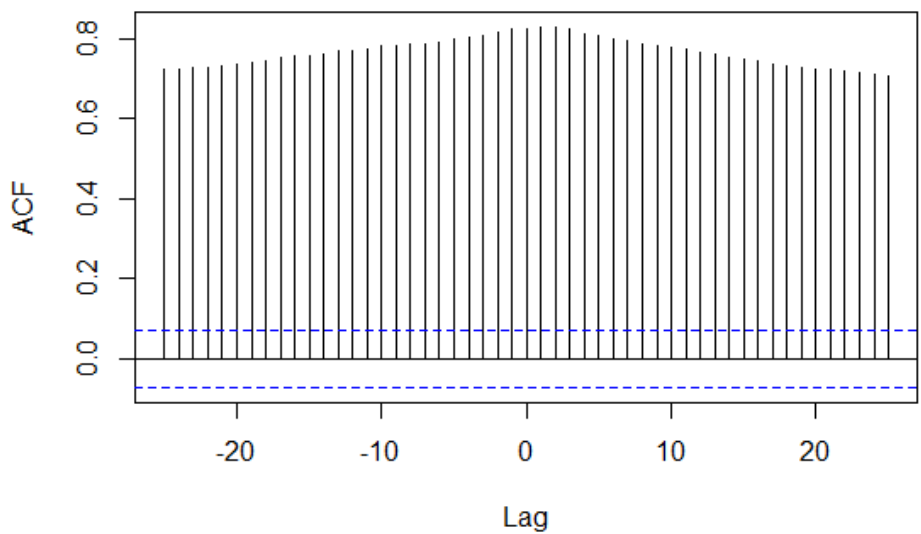

*Figure 4: Cross correlation of Perp prices with Spot prices, for a single simulation run with μ=1, σ=0.5,*

*S0=100, σ$_c$=10, σ$_n$=10. The highest correlation is at a lag of 2.*

### 4.1 Analysing the accuracy of the Peg

To analyze the properties of the 'peg', we use Shewhart Charts (Figure 3) - a line graph representing a measure over time, with bounds representing an upper and lower control limit typically set at ±3 standard deviations from the mean, referred to here as the 'Center'. This method is used extensively in control theory for processes that have an objective function to target, within acceptable tolerances. In the case of Perpetual Future Premiums, the Center approximates the Spot Price at time t, as the $Premium_t$ is the deviation of the Perp price from the spot price at time t. The points marked in red are premiums that are more than 3 standard deviations (represented by the Upper and Lower Control Limits, UCL and LCL lines in the chart below) away from the mean (the Control Line, CL, in the chart below), referred to as 'Violations'. The points in orange represent 'Runs' consecutive sequences of premiums that move away from the Center - an indication of the process deviating from the norm. These Shewhart Chart estimates will be our primary tools to explore the quality of the peg across a diverse range of parameters in the simulations below.

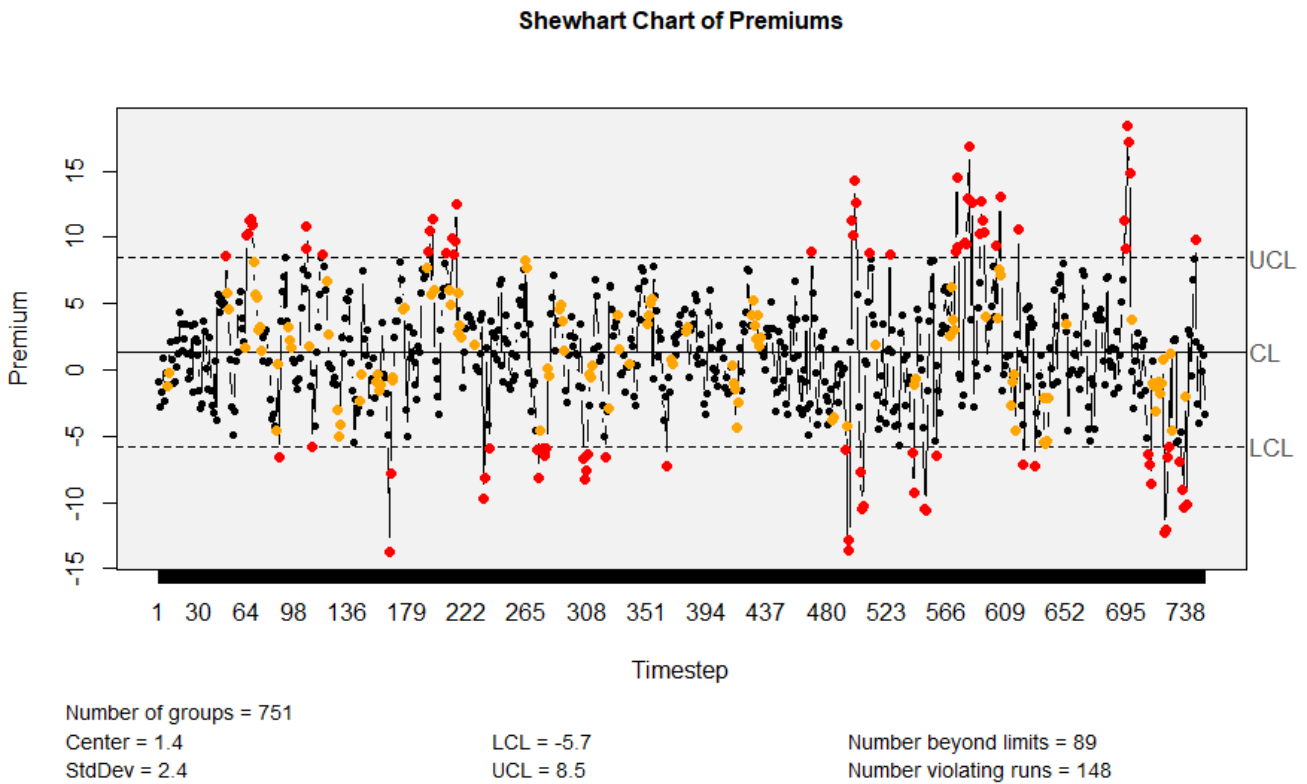

*Figure 5: The control chart for a simulation run with μ = 1, σ=0.5, S0=100, σ$_C$=10, σ$_n$=10. The upper and lower control limits are 8.5 and 5.7 respectively, with the center of the distribution at 1.4, slightly above the expected 00. The number of 'violations' - Premiums greater than 3 standard deviations from the center - are 89, marked in Red. Finally, the number violating runs is 148 - this is a count of the number of consecutive premiums that are continuously increasing or decreasing away from the center - an indication of the process deviating from the norm (in Orange).*

### 4.2 Simulation Parameters

The behavior of the traders is entirely determined by the two price signals from the Perp market itself, the Spot market, and the following parameters:

Forecasting Parameters:

$\sigma_f$: The weightage given by a trader to fundamentalist strategies. This is set to 0 as discussed in the methods section above.

$\sigma_c$: The weightage given by a trader to chartist strategies

$\sigma_n$: The weightage given by a trader to noise strategies

$\sigma_\epsilon$: A volatility parameter used by the noise strategy to produce forecasts

$l_{min}, l_{max}$: The time horizon used by the Chartist strategy to generate forecasts

Trading parameters:

$k_{max}$: The range above and below the forecast price from which the trader selects their Bid and Ask prices

$bias$: This parameter modulates the tendency for a trader to prefer positional or basis trading startegies based on whether they are trading long or short.

$exit_{probability}$: The probability of exiting the market at each timestep

$\tau$: Only the last $\tau$ Bids and Asks are retained in the orderbook at each timestep.

$cohort_{size}$: The number of randomly drawn traders who participate in trading at each timestep

### 4.3 Exploratory Simulations

We start by exploring the effect of varying the forecasting parameters on the 'peg' of the Perp price to the Spot price. This is done first to identify suitable relative weights of trader strategies, enabling us to fix them and thereafter examine the effects of the other parameters on Market dynamics. As discussed in the methods section, the fundamental strategy has zero weight, because it has no meaning in the Perpetual Futures context. This leaves the relative weights of Chartist and Noise traders to consider. By keeping the chartist weight constant at 10 and varying the Noise weight from 1 to 10, we can see a range of regimes where the Noise trader varies from $\frac{1}{10}$ of the Chartist trader to the Noise trader being equally weighted compared to the Chartist.

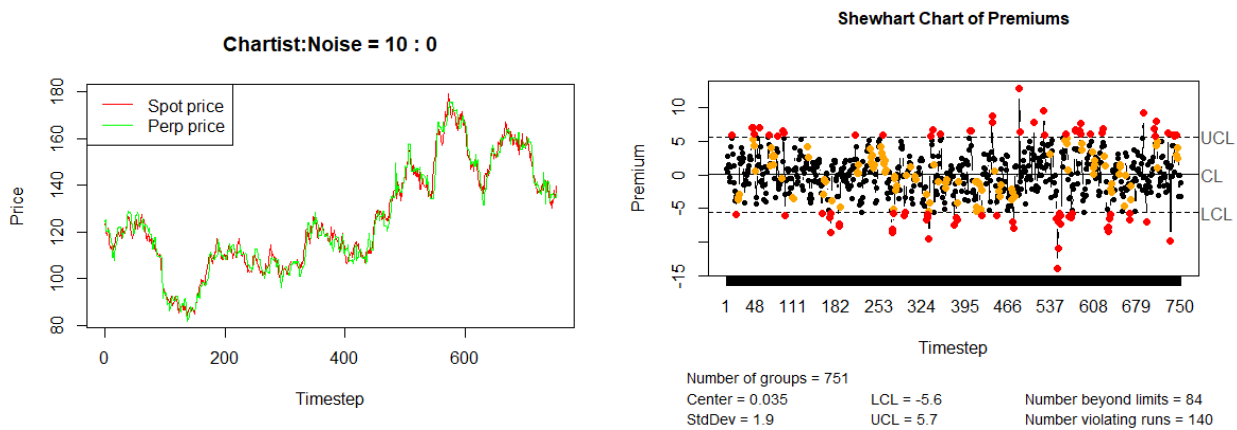

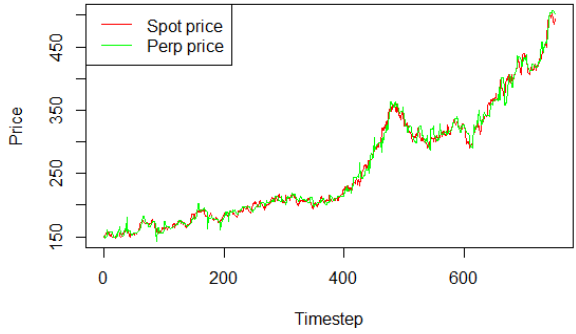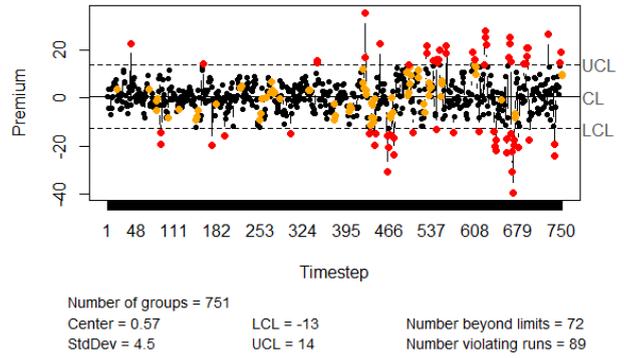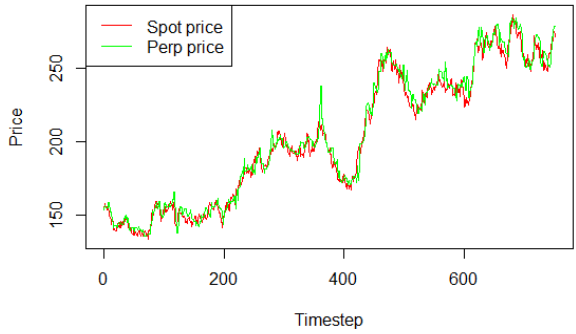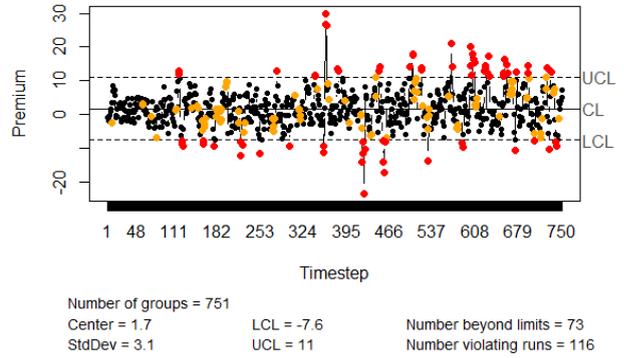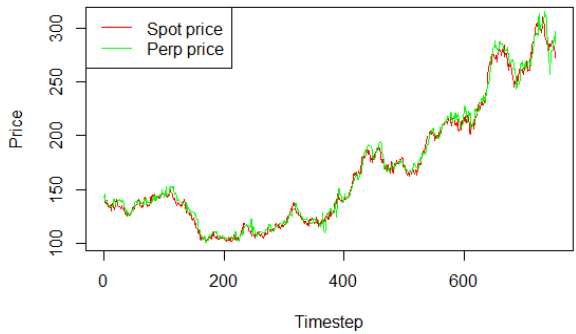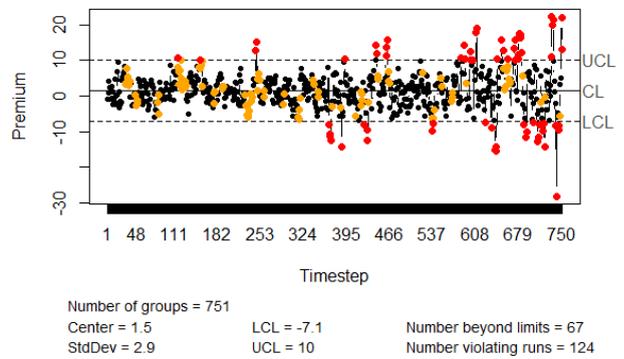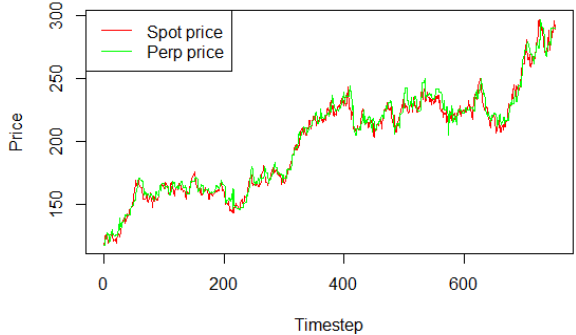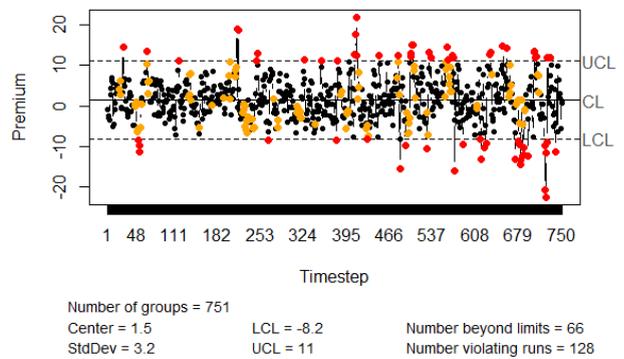

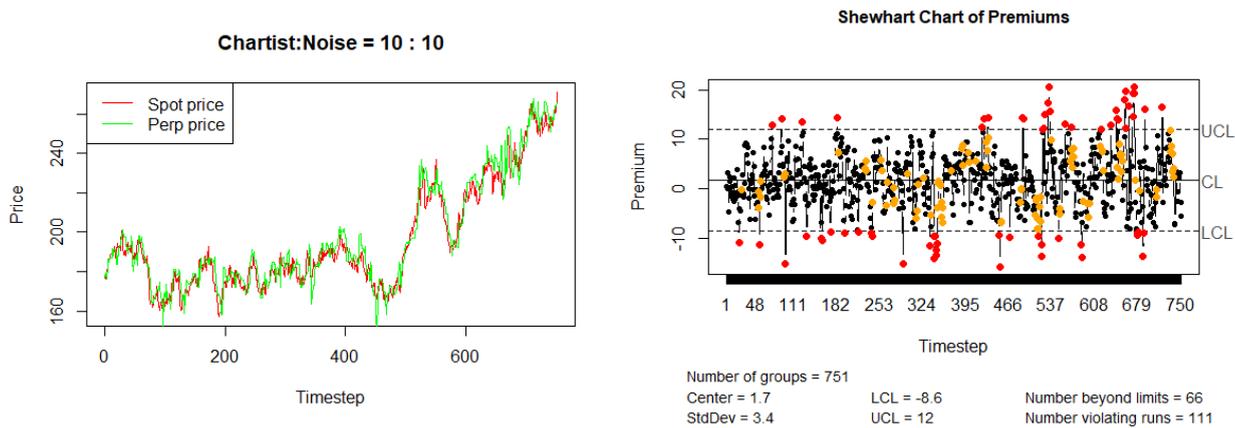

*Figure 6: Parameter sweep of the relative weights of Chartist vs Noise traders. Here we see the results of Chartist:Noise weights from 10:0 to 10:10 in increments of 2. The price 'peg' of the Perp to the underlying Spot is not very sensitive to the trading strategies adopted by traders in this market, suggesting the funding rate mechanism is effective in maintaining the price correlation in a wide range of market conditions*

Varying the relative weight of Chartist and Noise strategies does have some impact on the sensitivity of the peg, as evidenced by the topmost images in Figure 2, which displays the results of a simulation run with a 'Chartist-only' strategy. This strategy produced the tightest bounds of control at -5.6 and 5.7. The general trend of maintaining the peg is, however, present across the entire range of the runs, and does not vary greatly. This appears to echo the results of the original paper which found that varying the strategy weights produced results that remain 'qualitatively the same' (Chiarella et al, 2002). Further, with only noise traders present, we find that the peg appears to drift away from the spot price with significantly greater variation over time (Figure 5). This also echoes the findings of Chiarella et al. (2002), which suggests that the presence of both types of traders is necessary to produce more realistic simulations. It should be noted that this result is sensitive to $\sigma_\epsilon$, which is the volatility parameter used by the noise trader to produce random forecasts. Varying $\sigma_\epsilon$ does vary the error represented by the Premium but is not of primary interest in examining the properties of the simulation. This is because naturally increasing the error of the noise trader can be expected to introduce variability in the results and does not shed any light on the microstructure of a Perpetual Futures market. Also, all the simulation runs show an increase in 3-sigma violations over time, a feature that will not be examined in greater detail in this paper due to limitations of scope.

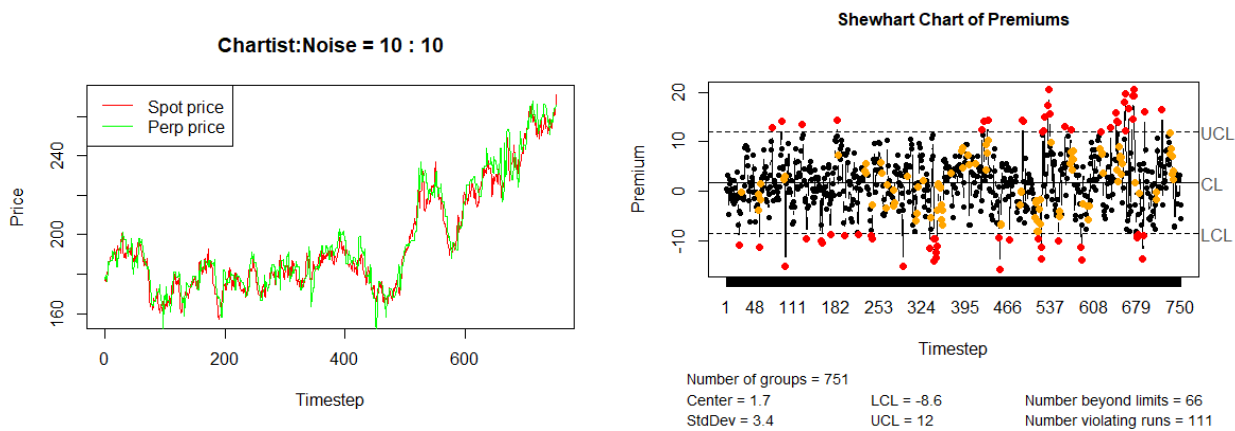

*Figure 7: If only Noise traders are present with a $\sigma_\epsilon$=0.05, the peg still holds, but with greater deviation than if Chartist traders are also present.*

This prima-facie, naive exploration suggests that to explore the market dynamics represented by the other parameters, we can utilize equal weights of Chartist and Noise traders to provide an unbiased basis for exploring the other parameters. The remaining simulation runs in the paper will feature equally weighted Chartist and Noise traders.

### 4.4 Simulations

The primary simulations explore the effect of varying the parameters $l_{max}$, $k_{max}$, $\tau$, and *bias*. The rest of the parameters were fixed at the following values:

$\mu$ (drift of GBM) = 1
$\sigma$ (volatility of GBM) = 0.5
$S_0$ (starting value of GBM) = 100
$\sigma_f$ (Fundamentalist weight) = 0
$\sigma_c$ (Chartist weight) = 10
$\sigma_n$ (Noise weight) = 10
$\sigma_\epsilon$ (Noise volatility parameter) = 0.05
$l_{min}$ (lower end of Time Horizon parameter) = 1
$exit_{probability}$ (market exit probability) = 0.05
$\tau$ (range of latest live orders) = 8
$cohort_{size}$ (number of traders who enter the market at each timestep) = 4

In the simulation runs below, a hundred simulation runs are made for each value of the parameter being varied, and the estimates of the Shewhart charts (Upper Control Limit, Lower Control Limit, Violations and Runs) are averaged across these 100 runs. These values are then plotted to see how the model responds to different values of these estimates.

### 4.5 Varying the upper bound of the Chartist Time Horizon Parameter $l_{max}$

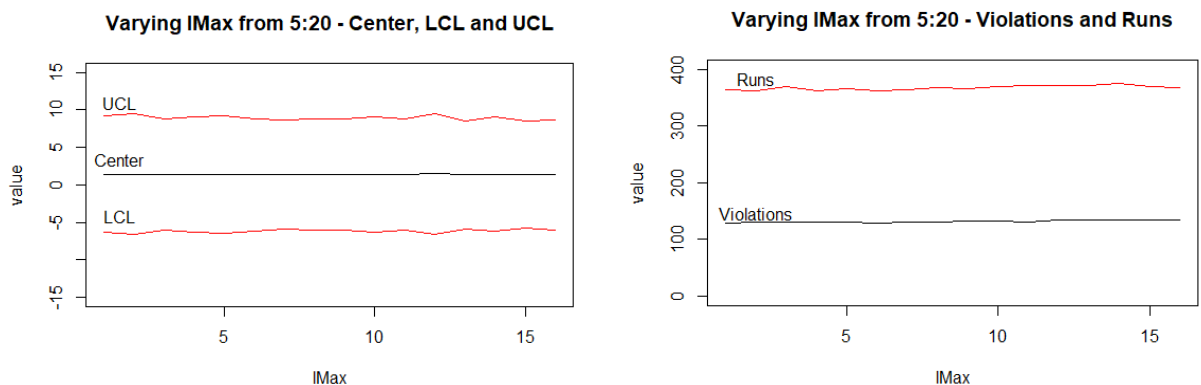

*Figure 8: On the left, we see that the Center, the Lower Control Limit (LCL), and the Upper Control Limit (UCL) are stable across a range of $l_{max}$ from 5 to 20, all other parameters being constant.*

100 simulations were run for each level of $l_{max}$, ranging from 5 to 20. The model is stable across the entire range of this parameter sweep of $l_{max}$ over the interval (5, 20), showing no trend across the sweep. This parameter controls how far into the past the Chartist trader strategy looks to generate a forecast for the price. At each assay the trader picks a horizon between $l_{min}$ and $l_{max}$ to generate a moving average and extrapolate the price for the next timestep. This suggests that the model is robust to a wide range of Chartist rime horizons.

## 4.6 $k_{max}$

$k_{max}$ controls the 'spread' of the Bids and Asks. At each assay, the trader first generates a composite price forecast, and then chooses a spread factor $k$ from the interval (0, $k_{max}$). If generating an Ask, the trader prices it by:

$$Price_{forecast} * (1 - k)$$

If producing a Bid, the trader prices it by:

$$Price_{forecast} * (1 + k)$$

Therefore, the lower $k$ is, the tighter the Spread of the Bids and Asks, and the more efficient the market.

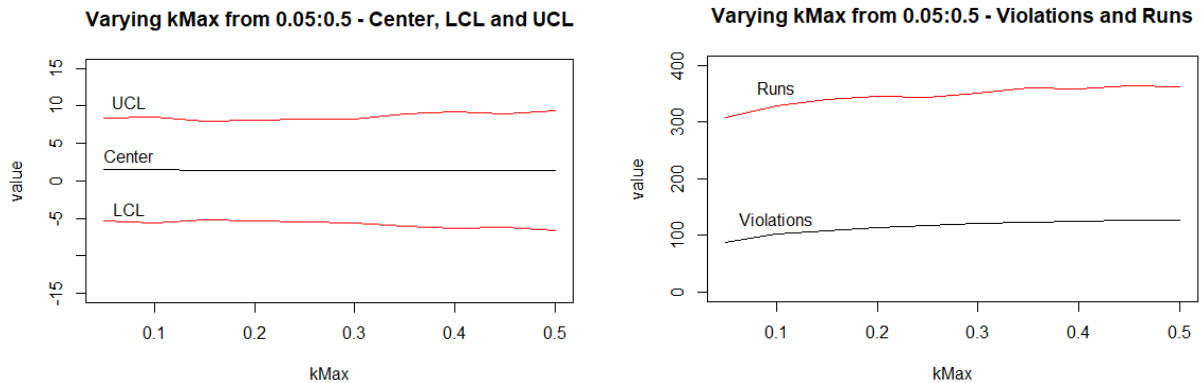

*Figure 9: On the left, we see that the Center, the Lower Control Limit (LCL), and the Upper Control Limit (UCL) are stable across a range of $l_{max}$ from 5 to 20, all other parameters being constant.*

This is borne out by the simulation runs, with a marginally lower UCL and LCL spread at lower values of $k_{max}$ and an increase in Runs and violations as $k_{max}$ increases.

## 4.7 τ

τ controls the depth of the orderbook in this simulation since a fixed cohort of traders always enters the market to trade at each timestep. After each assay, only the latest τ Bids and Asks are retained in the orderbook, and all older trades are removed. Therefore, the larger τ is, the more Bids and Asks are present in the orderbook, making it deeper.

Empirically, a deeper orderbook is expected to tighten the spread of Bids and Asks.

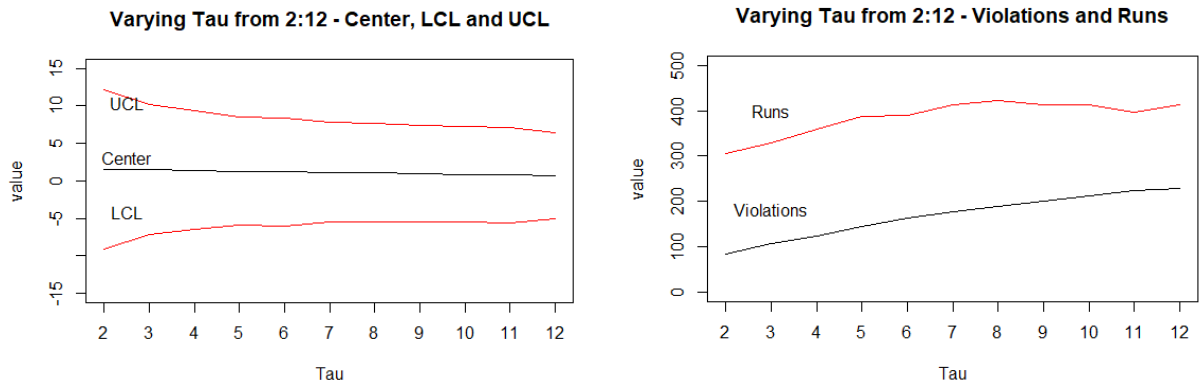

*Figure 10: On the left, we see that the Center, the Lower Control Limit (LCL), and the Upper Control Limit (UCL) are stable across a range of $l_{max}$ from 5 to 20, all other parameters being constant.*

This is borne out by the simulation, with the UCL and LCL spread tightening as τ increases. However, this comes with an associated increase in Violations and Runs which can be explained by the presence of stale orders from an earlier price regime, as well as the tighter control itself. This dynamic can be studied further with respect to empirical data from real Perp markets.

**4.8 Bias**

The *bias* parameter simultaneously determines the propensity of a Long trader to trade positionally and the Short trader's propensity to basis-trade. If the *bias* is 0.2, the Long traders are 80% likely to trade positionally, and the Short traders are 80% likely to basis-trade. At a bias of 0.5, both types of traders are equally likely to basis-trade or trade positionally. At values greater than 0.5, the propensities are reversed.

Empirically, it is observed that Longs tend to trade positionally, and Shorts tend to basis-trade, which results in Perps tending to trade at a premium to the Spot price. I varied the *bias* parameter over the interval $(0.05, 0.5)$ to see if the simulation can reproduce this nuanced phenomenon.

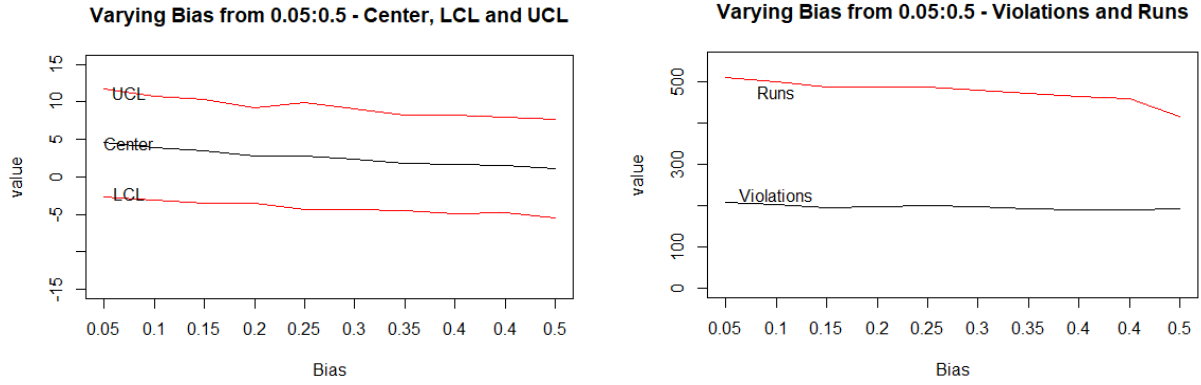

*Figure 11: On the left, we see that the Center, the Lower Control Limit (LCL), and the Upper Control Limit (UCL) are stable across a range of $l_{max}$ from 5 to 20, all other parameters being constant.*

The simulation bears out this phenomenon. At low levels of *bias* (Longs trade more positionally, shorts basis-trade more), the Perp market trades at a premium to the Spot market, with a center averaging near $+5$. As the bias goes closer to $0.5$, the Center tends towards 0. This is accompanied by a reduction in Runs, and marginally, Violations as well.

This result is interesting because the extremely simple nature of the agents and market design is sufficient to not only reproduce the pegging of the Perp price to the Spot price, but also reproduce a nuanced phenomenon which is observed empirically.

## 5. Conclusions

In this paper I introduce a novel Agent-based Model of a Perpetual Futures market. I show that simple canonical agents operating under the minimal market mechanisms described in the paper are sufficient to reproduce the salient features of a Perpetual Futures market. The focus is to explore if the model can reproduce the most salient feature of the Perp market, the pegging of the Perp price to the Spot price, an emergent effect of rational agents trading to maximize profits. This was achieved by extending a simple canonical model of a stock market to incorporate the richer features and context of the Perp market. In this process I introduce Long and Short positions as well as positional and basis-trading to the behaviors of the agents. To maintain the salient properties of the original work, this was done without changing the internals of agent behavior, but simply achieved by selecting the appropriate price signal (Spot prices for positional traders, Premia for basis-traders) that agents use to generate their forecasts. I introduced two parameters reflective of real-world behavior to the original model, $exit_{probability}$ and $cohort_{size}$, which modulate the number of traders who enter, and the rate at which they exit the market.

While this produced a rich environment with many potential avenues of exploration, the focus of my analysis was to investigate the properties of the 'Peg' in response to the dynamics of model parameters. This was done primarily using Shewhart charts of the Perp premiums (defined as the difference between the Perp price and the Spot price) which produce five point estimates that reflect different aspects of the Peg. The *Center* is the mean of the premiums, a measure of the average displacement of the Perp price from that of the Spot price. The LCL and UCL measure the variation around this *Center*, and reflect the 'accuracy' of the peg over time. Lastly, Violations and Runs reflect the counts of outliers in the Perp prices, departing far from the intended peg.

Decreasing the 'Spread' by varying the $k_{max}$ parameter resulted in lower numbers of Violations and Runs, while only marginally affecting the spread of the UCL and LCL.

Increasing the depth of the orderbook via τ resulted in a tighter spread of UCL and LCL, which is intuitive, while also increasing the numbers of Violations and Runs.

Lastly, varying *bias* to reflect the propensity of Longs to trade positionally and Shorts to basis-trade successfully reproduced a nuanced empirical phenomenon where this bias leads to the Perp market trading at a premium to the Spot market.

While the model and subsequent simulations runs were able to extend the original work to the more complex and nuanced Perp markets, some important avenues of exploration were left untouched.

In terms of design, the most significant of these is the fixing of liquidity in the market by making the $cohort_{size}$ a constant parameter. The effect of liquidity on market behavior is a very important and interesting dynamic (which was explored in the original paper via the λ parameter) which was not examined in this work. Similarly, tick size, which represents the trading frequency features of traders was not explored (explored in the original paper via the $\Delta\,parameter$). Agent wealth and dynamic order-sizing were also not explored, as was also the case in the original work.

In terms of analysis, there is a lot of potential to explore many other aspects of the Market apart from the single focus on the peg of Perp prices to Spot prices. Using a real Spot price signal instead of a generated signal and comparing the Perp prices to real Perp prices can help tune the parameters of the model to better reflect real markets. Returns, volatility, and other salient features were also not examined due to limitations of scope.

These limitations represent a rich field of potential enquiry into the properties of this model, and its relevance to real Perp markets. As Perps become the dominant arena of price discovery in cryptocurrency markets, such enquiries can lead to significant insight into the nature of these markets. My results suggest that the model is robust and analogous to real world markets, reproducing many features empirically observed. Perpetual Futures are a very important emerging derivative type with limited coverage in the literature. This work can represent a robust foundation upon which future enquiry can be developed.

# Appendix I

Complex Systems

#emergentproperties: In the introduction, I discuss the focal point of this thesis, the 'peg' of the Perp price to the Spot price through the 'Funding Rate' mechanism. This peg arises not out of an overt rule or directive to the trades, but rather as the consequence of maximizing their profits under the rules of the system. The primary objective of the thesis is indeed to show that the interactions of simple agents can reproduce this emergent effect. In the results section, I discuss the effect and interpretation of the effects of varying parameters on the point estimates of the Premia. The aggregated estimates plotted against the value of the parameter reveal complex dynamics that are once again the emergent effect of interactions between traders and the system. For example, I show how deepening the orderbook by increasing the parameter $\tau$ causes the upper and lower control limits to tighten – improving the efficiency of the system. The effect of Biasing traders is another example, where this causes the Perps to trade at an overall Premium to the Spot price – another emergent effect of the complex interplay of traders and the system.

#levelsofanalysis: In the literature review, I demarcate the field at multiple levels of analysis after introducing Perpetual Futures. Starting at the highest level of generality, I motivate the approach of Agent-based Computational Economics (ACE) at the level of the field. I identify and explain the properties of such approaches that provide insight where analytic models fail. I then move on the level of ACE itself and describe the evolution of the field in terms of the 'three stages of ACE', and place my work in that context. Finally, at the level of the simulation, I discuss the work of Chiarella et al. (2002) in terms of the particulars of the model and foreshadow the ways in which my own work builds on the original work. The choices made in this thesis is thereby motivated at multiple levels of analysis, starting at the highest level of generalization, and stepping down towards the relevant particulars.

Empirical Analyses

#modeling: A Perpetual Market is a complex system with emergent features (the peg) that are not easy to derive analytically from first principles. If a simple model can reproduce the observed phenomenon, the parameters of the model can be investigated for explanatory power. This is achieved successfully in this paper, with the peg reproduced in the simulated Perp Market with very minimal agents designed for a far simpler environment. Numerous choices were involved in achieving this without 'baking the result in' – including removing the Fundamentalist trader and retaining the original Agent behavior by simply switching the price signal that they use to generate forecasts based on whether they are trading positionally or basis-trading. Considering the necessity of exiting traders to prevent the possibility of skewed static populations of Long and Short traders tainting the results is another such example. Modelling is finally an exercise of creating an expressive parameter space that can be manipulated to perform experiments, which is seen in the results section where sweeps of parameters echo empirically observed phenomena such as the effect of spread, depth and bias on the Perp market.

#dataviz: Throughout the thesis, plots at the appropriate level of detail play the role of explaining complex concepts without requiring extensive verbal description. Examples of this are Figure 1, where the funding rate is explained, Figure 2, where positional and basis-trading are contrasted, the Shewhart charts used throughout, where the features of the seemingly messy Premia are extracted as the Center, UCL, LCL, Runs and Violations – enabling the numerical characterization of the rather fuzzy notion of 'a peg'. Finally, the results of the parameter sweeps portray almost simplistically the concepts of depth of the orderbook tightening the spread, and the bias causing the Perps to trade at a premium to the Spot market. All the figures in the Thesis aim to directly convey the intended and often complex message without requiring interpretation and cognitive labor.

Formal Analyses

#algorithms: To create this simulation, the complex behaviors and rules of the trading system had to be codified into a series of steps that would be iterated through the timesteps and multiple simulation runs to produce a convincing model of a Perp market. Expressing such a complex system simply and intelligibly is important as beyond replication, the algorithm has to be amenable to parametric manipulation to produce different effects. At the level of a single run, the algorithm had to be able to reproduce a single trading session. To produce a parameter sweep, each run itself becomes a component of a larger algorithm that varies the relevant parameter and averages across a hundred sessions. At each timestep, cohorts of traders would enter and produce trades according to a range of behaviors such as being Long, Short, trading Positionally or Basis-trading. Thus the simulation requires three levels of algorithmic design, while still being simple enough to reason about with respect to the real world system.

#descriptivestats: Identifying the language of Shewhart charts to characterize the rather amorphous concept of the peg enabled the exploration of the parameter sweeps to be straightforward and easy to interpret. This was essential, as the plot of Perps vs Spot, or that of the Premiums do not immediately convey the nuanced dynamics of the system. But observing the way in which the Center, UCL, LCL, Violations and Runs varied over different parameter sweeps reveals a wealth of information about the system. The plot of these estimates against different values of τ is one example, as is the plot against the variation of the bias. In both cases the message conveyed by these stats is immediately cognizable, without requiring complex analysis. A descriptive statistic should surface important insight directly, and in this paper I think it has been successful.

Multimodal Communications

#context: Perpetual Futures are an esoteric derivative type that may be unfamiliar to even technical audiences familiar with the finance industry. By describing its origin in traditional finance, then situating its implementation in the Crypto industry, and finally placing it in the context of ACE models, I cover the required understanding of this important innovation in multiple aspects necessary to the paper. Similarly, in the literature review, I place my work I the larger context of ACE, the three stages of ACE, few type models, and finally the work of Chiarella et al. whose work I extended. Finally, in the conclusion, I revisit the context to address the limitations of the current work, and provide concrete suggestions for addressing them and motivating future work.

#thesis: From the Conclusion: 'Simple canonical agents operating under the minimal market mechanisms described in the paper are sufficient to reproduce the salient features of a Perpetual Futures market'. This thesis statement lends explanatory power to the parametrization of the model and provides a robust foundation upon which to base future work. The ability of the parameter sweeps to convey complex ideas such as 'orderbook depth tightens the peg' is a direct consequence of the formulation of the primary goal of this paper. Throughout the paper, this thesis is reinforced in terms of measurable, relevant, and insightful results. Each of the results is itself precisely expresses an aspect of the thesis statement that contributes to the primary thesis without requiring supplementary analysis and explication, such as the result that skewing long traders to trade positionally, and short traders to basis-trade causes the Perp market to trade at an overall premium to the Spot market. I believe the paper is able to achieve this due to the overarching thesis acting as an organizing principle for the whole work.

# Appendix II

*All code can be accessed here: https://github.com/Ramshreyas/MinervaMDAThesis*

**Agent Definition**

Agent.R

```r
# Helper function to create agents as simple vectors of their parameters:
createAgentVector <- function(sigmaF, sigmaC, sigmaN, kMax, lMin, lMax) {
  
  c(sigmaF * runif(1),    # Fundamentalist weight
    sigmaC * runif(1),    # Chartist weight Randomized Spread size
    sigmaN * runif(1),    # Noise Weight
    sample(lMin:lMax, 1), # Horizon range for Chartist momentum rules
    kMax * runif(1),      # BId/Ask spread parameter
    -1)
}

# Helper to get Agent parameters
getAgentParameter <- function(agent, param) {
  agent[, param]
}

# Helper to set Agent parameters
setAgentParameter <- function(agent, param, value) {
  switch (
    param,
    "Fundamentalist" = { agent[1] <- value },
    "Chartist" = { agent[2] <- value },
    "Noise" = { agent[3] <- value },
    "Horizon" = { agent[4] <- value },
    "Spread" = { agent[5] <- value },
    "Side" = { agent[6] <- value }
  )
  
  #Return Agent
  agent
}
```

**Forecasts**

Forecast.R

```r
# Function that generates a forecast given a trader, the spot price history and the perp price history
getForecast <- function(trader, prices, sigmaE, t) {
  
  # The Fundamentalist believes the price reverts to the mean. The return expectation is scaled by the number of ticks
  fundamental_forecast <- log(mean(prices)/prices[t])
  
  # The chartist averages the sum of simple returns [P(t) - P(t-1)]/P(t-1) over a time horizon randomly chosen from [lMin, lMax]
  l <- sample(lMin:lMax, 1)
  horizon <- t - l
  simple_returns <- (prices[t:horizon] - prices[(t-1):(horizon - 1)])/prices[(t-1):(horizon - 1)]
  chartist_forecast <- (sum(simple_returns)/l)
  
  # The noise trader produces a random forecast
  noise_forecast <- sigmaE * runif(1)
  
  # Composite forecast
  fundamental_weight <- getAgentParameter(trader, "Fundamentalist")
  chartist_weight <- getAgentParameter(trader, "Chartist")
  noise_weight <- getAgentParameter(trader, "Noise")
  total_weight <- fundamental_weight + chartist_weight + noise_weight
  
  # The combined weighted expectation
  price_expectation <- (fundamental_weight*fundamental_forecast + chartist_weight*chartist_forecast + noise_weight*noise_forecast)/total_weight
  
  # Return the forecast
  forecast <- prices[t]*exp(price_expectation)
  forecast
  
}
```

Orderbook functions

orderbook.R

```r
# Get Agent Order
getOrder <- function(trader, premia, prices, t, bias, close_position_probability, sigmaE) {
  # Get agent params
  k <- runif(1, 0, getAgentParameter(trader, "Spread"))

  # Randomly close positions for some traders and exit (allowing for trader distribution to be dynamic over time)
  if (runif(1) < close_position_probability) {
    setAgentParameter(trader, "Side", -1)
    #print("Exiting")
    return(list(trader, 0, 0))
  }

  # Generate price forecast
  price_forecast <- getForecast(trader, prices, sigmaE, t)

  # Get trader side
  side <- getAgentParameter(trader, "Side")

  # If new trader, randomly assign a side
  if (side == -1) {
    trader <- setAgentParameter(trader, "Side", sample(c(0,1), 1))
    #print("Randomly assigning side")
  }

  # If long trader
  if (getAgentParameter(trader, "Side") == 0) {
    # Assign basis or positional trade based on bias
    if (runif(1) > bias) {
      #print("Long trader positional")
      # Generate and return positional order
      return(getPositionalOrder(price_forecast, prices[t], trader, k))

    } else {
      #print("Long trader funding")
      # Adjust premia by adding lowest negative value to make all data positive
      adjusted_premia <- premia + abs(min(premia[1:t])) + 1

      # Generate funding forecast
      forecast <- getForecast(trader, adjusted_premia, sigmaE, t)

      # Generate and return order
      return(getFundingOrder(forecast, price_forecast, adjusted_premia[t], trader, k))
```

```r
      }

      # Else short trader
    } else {
      # Assign basis or arbitrage trade based on bias
      if (runif(1) < bias) {
        #print("Short trader positional")
        # Generate and return short positional order
        return(getPositionalOrder(price_forecast, prices[t], trader, k))
      } else {
        #print("Short trader funding")
        # Adjust premia by adding lowest negative value to make all data positive
        adjusted_premia <- premia + abs(min(premia[1:t])) + 1

        # Generate funding forecast
        forecast <- getForecast(trader, adjusted_premia, sigmaE, t)

        # Generate and return order
        return(getFundingOrder(forecast, price_forecast, adjusted_premia[t], trader, k))
      }
    }
}

# Helpers
getPositionalOrder <- function(forecast, price, trader, kMax) {

  # Get side of market
  side <- getAgentParameter(trader, "Side")

  if (forecast > price & side == 0 | forecast < price & side == 1) {
    #print("BUY")
    # Return Buy order
    return(list(trader, forecast*(1-kMax), 1))
  } else {
    #print("SELL")
    # Return Sell order
    return(list(trader, -1*forecast*(1+kMax), 1))
  }
}

getFundingOrder <- function(forecast, price, premium, trader, kMax) {

  # Get side of market
  side <- getAgentParameter(trader, "Side")
```

```r
    if (forecast < premium & side == 0 | forecast > premium & side == 1) {
      #print("BUY")
      # Return Buy order
      return(list(trader, abs(price)*(1-kMax), 1))
    } else {
      #print("SELL")
      # Return Sell order
      return(list(trader, -1*abs(price)*(1+kMax), 1))
    }
}

addOrder <- function(ob, order, t) {
  price <- order[[2]]
  size <- order[[3]]
  # print("Price:")
  # print(price)

  # Buy
  if(price > 0) {

    # Get best ask
    best_ask <- best.ask(ob)[["price"]]

    # If price less than current best ask
    if (is.na(best_ask) | price < best_ask) {

      # Set limit order
      ob <- add.order(ob, price, size, type = "BID", time = t, id = t)
      return(list(ob, NULL))

    } else {

      # Else add market order
      ob <- market.order(ob, size, "BUY")
      return(list(ob, best_ask))

    }
  # Sell
  } else if(price < 0) {

    # Remove negative sign for sell
    price <- abs(price)

    # Get best bid
    best_bid <- best.bid(ob)[["price"]]
```

```r
    # If price greater than current best bid
    if (is.na(best_bid) | price > best_bid) {

      # Set limit order
      ob <- add.order(ob, price, size, type = "ASK", time = t, id = t)
      return(list(ob, NULL))

    } else {

      # Else add market order
      ob <- market.order(ob, size, "SELL")
      return(list(ob, best_bid))

    }
    # Exit market
  } else {

    return(list(ob, NULL))

  }
}

removeOldOrders <- function(ob, tau, t) {
  if(t > tau) {
    for(i in 1:t - tau) {
      # If so, remove
      ob <- remove.order(ob, i)
    }
  }

  ob
}
```

Sample Simulation Run

debug.R

```r
source("forecast.R")
source("agent.R")
```

```r
source("orderbook.R")

stats <- list()
count <- 1

for(bias in c(0.05, 0.1, 0.15, 0.2, 0.25, 0.30, 0.35, 0.4, 0.4, 0.5)){

nSims = 100

perp_sims = data.frame(matrix(nrow = 1001, ncol = nSims))
spot_sims = data.frame(matrix(nrow = 1001, ncol = nSims))
prem_sims = data.frame(matrix(nrow = 1001, ncol = nSims))

for(n in 1:nSims) {

tMax <- 1000

spot_price <- gbm(x0=100, mu=1, sigma=0.5, t0=0, t=1, n=tMax)

# Set number of agents
nAgents <- 200

# Fundamentalist weight
sigmaF <- 0

# Chartist weight
sigmaC <- 10

# Noise weight
sigmaN <- 10

# random component of spread
kMax <- 0.5

# horizons for momentum rules - bounds for how far back should they go to estimate trend
lMin <- 1
lMax <- 5

# Initialize traders list
traders <- data.frame(matrix(ncol = 6, nrow = 0))
names(traders) <- c("Fundamentalist", "Chartist", "Noise", "Horizon", "Spread", "Side")

#bias <- 0.5

close_position_probability <- 0.05
```

```r
sigmaE <- 0.05

t <- 250

tau <- 8

cohortSize <- 4

perp_prices <- spot_price + runif(tMax+1, -1, 1)
perp_prices[250:tMax + 1] <- rep(0,tMax + 1)

premia <- perp_prices - spot_price
premia[250:tMax + 1] <- rep(0,tMax + 1)

# Create traders
for (i in 1:nAgents) {
  traders[i,] <- createAgentVector(sigmaF, sigmaC, sigmaN, kMax, lMin, lMax)
}

ob <- orderbook("orderbook.txt")

for(i in (t - 2*tau):(t-1)) {
  if(i %% 2 == 0) {
    ob <- add.order(ob, spot_price[i] - sample(0:10, 1), 1, type = "BID", time = t, id = i)
  } else {
    ob <- add.order(ob, spot_price[i] + sample(0:10, 1), 1, type = "ASK", time = t, id = i)
  }
}

for (t in 250:tMax) {
  # Select random traders
  cohort <- traders[sample(1:nAgents, cohortSize), ]

  newPrice <- NULL

  # Iteratively make bids and asks on the same price point, exit if trade occurs
  for (row in 1:cohortSize) {
    # Get trader
    trader <- cohort[row, ]

    # Get order
      order <- getOrder(trader, premia, spot_price, t = t, bias, close_position_probability, sigmaE = sigmaE)
      price <- order[[2]]
```

```r
    size <- order[[3]]

    if(price == 0) next

    # Add to book as market or limit
    result <- addOrder(ob, order, t)
    ob <- result[[1]]
    newPrice <- result[[2]]

    # Check if trade occurs
    if (!is.null(newPrice)) {
      break()
    }
  }

  # If no trade occurred, set price to mid point
  if (is.null(newPrice)) {
    # Update perp price as midpoint between best ask and best bid
    newPrice <- mid.point(ob)
  }

  if(is.na(newPrice)) {
    #newPrice <- spot_price[t] + runif(1)
    newPrice <- perp_prices[t]
  }

  # Update perp_prices
  perp_prices[t+1] <- newPrice

  # Update premia
  premia[t+1] <- perp_prices[t+1] - spot_price[t+1]

  # Clean old trades
  ob <- removeOldOrders(ob, tau, t)
}

perp_sims[,n] <- perp_prices
spot_sims[,n] <- spot_price
prem_sims[,n] <- premia

}

p <- premia[250:tMax]

title <- paste0("Chartist:Noise = ", sigmaC, " : ", sigmaN)
```

```r
plot(spot_price[250:tMax], type = "l", col = "red", main = title, xlab = "Timestep", ylab = "Price")
lines(perp_prices[250:tMax], col = "green")
legend("topleft", legend=c("Spot price", "Perp price"),
       col=c("red", "green"), lty=c(1,1), cex=1)

qcc(data = p,
    type = "xbar.one",
    title = "Shewhart Chart of Premiums", # Replacement title
    xlab = "Timestep",
    ylab = "Premium",
    digits = 2, # Limit the significant figures
    plot = TRUE)

print("Sims done")

qs <- data.frame(matrix(nrow = 100, ncol = 6))
names(qs) <- c("center", "stddev", "lcl", "ucl", "violations", "runs")

for (i in 1:100) {
  q <- qcc(prem_sims[,i], type = "xbar.one", digits = 2, plot = FALSE)

  center <- q$center
  stddev <- q$std.dev
  lcl <- q$limits[1]
  ucl <- q$limits[2]
  violations <- length(q$violations[[1]])
  runs <- length(q$violations[[2]])
  qs[i,] <- c(center, stddev, lcl, ucl, violations, runs)
}

stats[[count]] <- qs

count <- count + 1

}
```

Data Plots:

debug.R

```r
centers <- stddevs <- lcls <- ucls <- violations <- runs <- c()

for(s in 1:(count-1)) {
  centers <- append(centers, mean(stats[[s]]$center))
```

```r
    stddevs <- append(stddevs, mean(stats[[s]]$stddev))
    lcls <- append(lcls, mean(stats[[s]]$lcl))
    ucls <- append(ucls, mean(stats[[s]]$ucl))
    violations <- append(violations, mean(stats[[s]]$violations))
    runs <- append(runs, mean(stats[[s]]$runs))
}

plot(centers, type = 'l', ylim = c(-15,15),
    main = 'Varying Bias from 0.05:0.5 - Center, LCL and UCL',
    xlab = 'Bias',
    ylab = 'value',
    xaxt = 'n')
axis(1, at = 1:10, labels = c(0.05, 0.1, 0.15, 0.2, 0.25, 0.30, 0.35, 0.4, 0.4, 0.5))
text(1.5, mean(centers) + 2, "Center")
lines(lcls, col = 'red')
text(1.5, mean(lcls) + 2, "LCL")
lines(ucls, col = 'red')
text(1.5, mean(ucls) + 2, "UCL")

plot(violations, type = 'l', ylim = c(0, 550),
    main = 'Varying Bias from 0.05:0.5 - Violations and Runs',
    xlab = 'Bias',
    ylab = 'value',
    xaxt = 'n')
axis(1, at = 1:10, labels = c(0.05, 0.1, 0.15, 0.2, 0.25, 0.30, 0.35, 0.4, 0.4, 0.5))
text(2, mean(violations) + 30, "Violations")
lines(runs, col = 'red')
text(2, mean(runs), "Runs")
```